\documentclass[useAMS,usenatbib]{mn2e}
\usepackage{graphicx,txfonts}

\newcommand{\cms}{\hbox{${\rm cm\,s}^{-1}$}}
\newcommand{\ms}{\hbox{${\rm m\,s}^{-1}$}}
\newcommand{\kms}{\hbox{${\rm km\,s}^{-1}$}}

\newcommand{\bspsmall}{\vspace{0.5cm}\small\noindent This paper has been typeset
from a \TeX/\LaTeX\ file prepared by the author.\normalsize}

\title[Frequency comb wavelength calibration]{High-precision
  wavelength calibration of astronomical spectrographs with laser
  frequency combs}

\author[M. T. Murphy et al.]{M. T.  Murphy$^{1}$\thanks{E-mail:
    mim@ast.cam.ac.uk (MTM)}, Th. Udem$^{2}$, R. Holzwarth$^{2}$, A.
  Sizmann$^{2}$, L. Pasquini$^{3}$,\newauthor C. Araujo-Hauck$^{3}$,
  H. Dekker$^{3}$, S.~D'Odorico$^{3}$, M. Fischer$^{2}$, T. W.
  H\"{a}nsch$^{2}$,\newauthor A. Manescau$^{3}$\\
  $^{1}$Institute of Astronomy, University of Cambridge, Madingley Road, Cambridge, CB3 0HA, UK\\
  $^{2}$Max-Planck Institut f\"ur Quantenoptik, Hans-Kopfermann-Strasse 1, 85748 Garching, Germany\\
  $^{3}$European Southern Observatory, Karl-Schwarzschild-Str.~2,
  85748 Garching, Germany}

\begin{document}

\date{Accepted ---. Received ---; in original form ---}

\pagerange{\pageref{firstpage}--\pageref{lastpage}}

\pubyear{2007}

\maketitle

\label{firstpage}

\begin{abstract}
  We describe a possible new technique for precise wavelength
  calibration of high-resolution astronomical spectrographs using
  femtosecond-pulsed mode-locked lasers controlled by stable
  oscillators such as atomic clocks. Such `frequency combs' provide a
  series of narrow modes which are uniformly spaced according to the
  laser's pulse repetition rate and whose absolute frequencies are
  known {\it a priori} with relative precision better than $10^{-12}$.
  Simulations of frequency comb spectra show that the photon-limited
  wavelength calibration precision achievable with existing echelle
  spectrographs should be $\sim$$1\,\cms$ when integrated over a
  $4000\,$\AA\ range. Moreover, comb spectra may be used to accurately
  characterise distortions of the wavelength scale introduced by the
  spectrograph and detector system. The simulations show that
  frequency combs with pulse repetition rates of $5$--$30$\,GHz are
  required, given the typical resolving power of existing and possible
  future echelle spectrographs. Achieving such high repetition rates,
  together with the desire to produce all comb modes with uniform
  intensity over the entire optical range, represent the only
  significant challenges in the design of a practical system.
  Frequency comb systems may remove wavelength calibration
  uncertainties from all practical spectroscopic experiments, even
  those combining data from different telescopes over many decades.
\end{abstract}


\section{Introduction}\label{sec:intro}

Echelle spectrographs are the basic tool for astronomical spectroscopy
with high velocity precision. Since many echelle diffraction orders
can be cross-dispersed and thereby recorded simultaneously on
rectangular-format media [e.g.~modern charge-coupled devices (CCDs)],
more-or-less continuous wavelength coverage over much of the optical
range can be achieved whilst maintaining high resolving
power\footnote{Resolving power $R \equiv \lambda/{\rm FWHM}$, where
  ${\rm FWHM}$ is the instrumental profile's full-width at
  half-maximum.}. Most modern echelle spectrographs operate at
$R=30000$--$150000$; higher resolving powers usually entail reduced
wavelength coverage. However, to fully exploit the velocity precision
available in high-resolution spectra, precise wavelength calibration
over much of the available wavelength range is usually required. For
example, searches for extra-solar planets via the radial velocity
method \citep[e.g.][]{MayorM_95a,MarcyG_96a} have achieved
night-to-night relative precisions better than $1\,\ms$ from stellar
absorption lines integrated over wavelength ranges of
$\sim$$1000$--$3000$-\AA\ \citep[e.g.][]{RupprechtG_04a,LovisC_06a}.
Also, echelle spectra of quasar absorption systems have constrained
cosmological variations in fundamental constants, such as the
fine-structure constant (e.g.~\citealt*{BahcallJ_67b};
\citealt{WebbJ_99a}) and proton-to-electron mass ratio
\citep[e.g.][]{VarshalovichD_93b,IvanchikA_05a}. Recent measurements
of $\alpha$-variation using many absorption systems reach relative
precisions of $\sim$$10^{-6}$ when comparing several transitions
separated by up to $\sim$$5000$\,\AA\ (e.g.~\citealt{MurphyM_01a};
\citealt*{MurphyM_03a}). The corresponding velocity precision required
is $\sim$$20\,\ms$.

The need for even higher precision is already clear. Detecting
Earth-mass extra-solar planets around solar-mass stars will require
$\la$$10\,\cms$ precision stable over many-year timescales. Also,
current evidence for varying constants
\citep[e.g.][]{MurphyM_04a,ReinholdE_06a} remains controversial and
such potentially fundamental results should be refuted or confirmed
with the highest possible confidence. There is even the possibility of
measuring the acceleration of the Universal expansion by monitoring
the change in redshift of Lyman-$\alpha$ forest absorption lines in
quasar spectra \citep[e.g.][]{SandageA_62a,LoebA_98a}. This would
require two epochs of spectra, taken decades apart, with velocity
precision approaching $1\,\cms$ integrated over
$\sim$$1000$--$3000$\,\AA\ ranges.  This is one possible aim of the
Cosmic Dynamics Experiment (CODEX) spectrograph proposed for the
European Extremely Large Telescope (\citealt{PasquiniL_06a};
\citealt{GrazianA_07a}; Liske et al.~2007, in preparation).

Astronomical echelle spectra are usually wavelength calibrated using
comparison emission- or absorption-line spectra. The ideal comparison
spectrum would comprise a series of lines which: (i) have known
wavelengths which are determined by fundamental physics; (ii) are
individually unresolved; (iii) are resolved from each other; (iv) have
uniform spacing; (v) cover the entire optical wavelength range; (vi)
have uniform strength (intensity or optical depth); (vii) are stable
over long time-scales and (viii) do not reduce the signal-to-noise
ratio ($S/N$) of the object spectrum. The calibration source should
also be: (ix) exchangeable, in the sense that two independent sources
produce the same spectrum; (x) easy to use, i.e.~essentially
`turn-key' operation for non-expert users and (xi) should have a
reasonably low cost. Table \ref{tab:req} summarises whether four
different calibration sources meet these requirements. The most
commonly used calibration source is the thorium-argon (ThAr)
hollow-cathode emission-line lamp. While it is relatively low cost,
has lines whose wavelengths are determined by atomic physics and which
cover the entire optical wavelength range, it has several
disadvantages. For example, ThAr lines differ widely in intensity and
spacing. Indeed, most are blended together at $R\la\!150000$.
Furthermore, as the current and pressure in the lamp varies (e.g.~as
the lamp ages) the relative intensity of different ThAr lines varies
widely. Iodine cells are also commonly used in radial-velocity
extra-solar planet searches \citep[e.g.][]{MarcyG_92a}. These have
similar advantages and disadvantages as ThAr lamps but they are
typically only useful over the wavelength range $5000$--$6500$\,\AA\
and absorb about half the object light, thereby significantly reducing
the $S/N$.

\begin{table}
\begin{center}
  \caption{The requirements for an ideal wavelength calibration source
    and the relative advantages and disadvantages of three current
    sources plus the frequency comb possibility we describe in this
    paper. We consider a spectrograph with resolving power
    $R\sim150000$.  `Poss.' means possible depending on the design and
    construction of the calibration source.}
\label{tab:req}
\begin{tabular}{lcccc}\hline
\multicolumn{1}{c}{Calibration lines} & \multicolumn{1}{c}{ThAr} & \multicolumn{1}{c}{I$_2$ cell} & \multicolumn{1}{c}{Etalon} & \multicolumn{1}{c}{Comb}\\\hline
From fundamental physics      & Yes       & Yes       & No       & Yes     \\
Individually unresolved       & Mostly    & Yes       & Poss.    & Yes     \\
Resolved from each other      & No        & No        & Poss.    & Poss.   \\
Uniformly spaced              & No        & No        & Yes      & Yes     \\
Cover optical range           & Yes       & No        & No       & Poss.   \\
Uniform strength              & No        & No        & Poss.    & Poss.   \\
Long-term stability           & No        & Poss.     & No       & Yes     \\
Maintain object $S/N$         & Yes       & No        & Yes      & Yes     \\
Exchangeable                  & Yes       & Yes       & Poss.    & Yes     \\
Easy to use                   & Yes       & Yes       & Poss.    & Poss.   \\
Reasonably low cost           & Yes       & Yes       & Yes      & Poss.   \\\hline
\end{tabular}
\end{center}
\end{table}

A significant step towards the ideal calibration source might be
achieved with the relatively new technology of optical `frequency
combs' generated from mode-locked femtosecond-pulsed lasers.
The 2005 Nobel Prize in physics was awarded
for the pioneering work on such combs (e.g.~\citealt{ReichertJ_99a};
\citealt{JonesD_00a}; \citealt*{UdemT_02a}). As the name suggests, a
frequency comb provides a spectrum of continuous-wave laser lines
called modes. They are produced by the repetitive pulse train of the
mode-locked laser. The mode spacing, which is constant in frequency
space, is given by the pulse repetition frequency. Since it resides in
the radio frequency domain it can readily be synchronised with an
extremely precise frequency reference such as an atomic clock.  The
comb's perfectly regular frequency grid should also facilitate the
effective compensation of non-linearities in the spectrograph's
calibration curve while directly linking the calibration to the
fundamental physics of the atomic clock. This direct link to the
definition of time and the long-term stability mean that even if
different combs are used to calibrate spectra taken at (perhaps
widely) different epochs, those spectra can be directly compared.
Research utilising archival and `virtual observatory' data would
directly benefit from such a reproducible calibration source.  These
properties would greatly facilitate extra-solar planet detection and
varying-constant analyses and would be more-or-less required for
detecting the Universal acceleration via Lyman-$\alpha$ forest spectra
taken decades apart.

This paper describes further the possibility of frequency comb
calibration of echelle spectrographs. Section \ref{sec:op} outlines
the basics of frequency comb generation, the methods used to ensure
that they are stable and the general manner in which they might be
used as practical echelle calibration sources. Section \ref{sec:phot}
derives the optimal laser repetition frequency and the overall
precision of the wavelength calibration that might be achieved with
typical echelle spectrographs. Some likely sources of systematic
errors which may otherwise limit the precision are also discussed.
Section \ref{sec:prob} outlines the main challenges in designing a
working frequency comb calibration system and proposes some possible
solutions.

\section{Basic operation}\label{sec:op}

\subsection{Generating the comb}\label{ssec:comb}

The exact properties of a frequency comb may be derived even without
knowledge of the operational principles of a mode-locked laser. It
suffices to know that these lasers can store a single pulse and
maintain it on a repetitive path. After each round-trip a copy of this
pulse is emitted though the output mirror/coupler of the laser,
resulting in an indefinite train of laser pulses. Like in any other
laser, the energy lost is replenished by stimulated emission in the
lasing medium, thereby allowing the energy in a short pulse to be
stored for any length of time by compensation of loss with gain.

\begin{figure}
\centerline{\includegraphics[width=0.9\columnwidth]{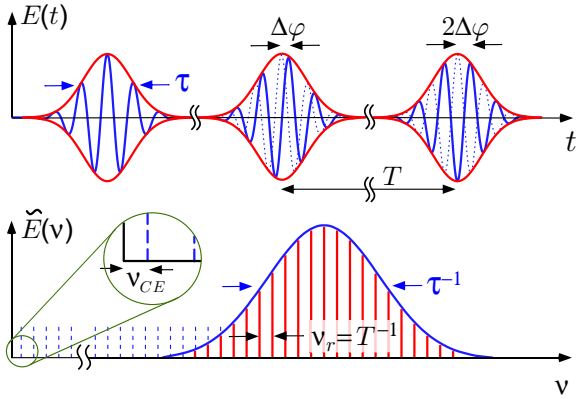}}
\vspace{-2mm}
\caption{A pulse train (top) produces the optical frequency comb in
  Fourier space (bottom) with some properties inverted. Shorter pulse
  envelopes produce broader frequency combs. The round-trip time of
  the pulses inside the generating laser cavity (not shown) determines
  the repetition frequency, $\nu_{\rm r}$, exactly. Dispersive
  elements in the laser cause a difference in the group and phase
  velocity, continuously shifting the carrier wave with respect to the
  envelope in the upper trace by $\Delta \varphi$ per pulse. In the
  frequency domain this causes the comb to shift by an amount
  $\nu_{\rm ce}=\Delta \varphi/2 \pi T$.}
\label{fig:pulse}
\end{figure}

To understand the mode structure of a femtosecond frequency comb and
the techniques employed to stabilize it, consider the idealized case
of a pulse circulating as a carrier wave with frequency $\nu_{\rm c}$
in a laser cavity with length $L$ . The cavity strongly modulates the
carrier wave amplitude which can therefore be described by a periodic
envelope function $A(t)=A(t-T)$, where $T$ is the round-trip travel
time of the pulse calculated from the cavity mean group velocity:
$T=2L/{\rm v}_{\rm gr}$. The repetition rate of the laser is simply
$\nu_{\rm r}=1/T$.  This is illustrated in Fig.~\ref{fig:pulse}. Since
the envelope function is periodic it can be written as a Fourier
series and the electric field at any given place (e.g.~the laser's
output mirror/coupler) can therefore be written as
\begin{equation}\label{eq:fourier}
 E(t) = A(t) e^{-i\,2\pi\nu_{\rm c}t}
      =  \sum_m A_m e^{-i\,2\pi(m\nu_{\rm r}+\nu_{\rm c})\,t}\,,
\end{equation}
where $A_m$ is the $m$th Fourier component of $A(t)$. Equation
(\ref{eq:fourier}) shows that, under the assumption of a periodic
pulse envelope, the resulting spectrum is a comb of laser frequencies
with spacing $\nu_{\rm r}$ such that the $m$th frequency component is
$\nu_m = m \nu_{\rm r} + \nu_{\rm c}$. In general, $\nu_{\rm c}$ is
not an integer multiple of $\nu_{\rm r}$ and so the modes are shifted
from being exact harmonics of $\nu_{\rm r}$. Re-numbering the modes
gives
\begin{equation}\label{eq:fopt}
  \nu_n =  n \nu_{\rm r} + \nu_{\rm ce} \hspace{1em}
\end{equation}
for $n$ a large integer, where the carrier envelope offset frequency,
$\nu_{\rm ce}$, is restricted to $0\leq\nu_{\rm ce}<\nu_{\rm r}$. This
equation maps two radio frequencies, $\nu_{\rm r}$ and $\nu_{\rm ce}$
-- which can be stabilized by a precise reference such as an atomic
clock -- onto the optical frequencies $\nu_n$.

Given the simple treatment above, one might expect deviations from
equation (\ref{eq:fopt}) in reality. However, mode locking the laser
ensures that equation (\ref{eq:fopt}) is realised to very high
precision in practice. The frequencies $\nu_n$ in equation
(\ref{eq:fopt}) represent the longitudinal cavity modes of the
mode-locked laser. The process of mode locking stabilizes the mode
frequency difference, $\nu_{\rm r}$, and the cavity mode phases such
that their superposition yields a short pulse. The existence of such a
pulse depends on the stability of the relative phases of the modes: if
these change in time -- for example, due to a small deviation of the
mutual mode separation from $\nu_{\rm r}$ -- the pulse would
disintegrate immediately. The very observation that the pulse stays
intact for a time $\Delta t$ implies that variations in $\nu_{\rm r}$
must be $<1/\Delta t$. For a pulse life-time of $1{\rm \,s}$, this is
only a few parts in $10^{14}$ at the optical frequencies $\nu_n$. Of
course, the stored pulse lasts much longer than $1{\rm \,s}$. In fact,
deviations from equation (\ref{eq:fopt}) -- that is, deviations from a
constant mode spacing -- have so far evaded detection even at the
$\sim$$10^{-16}$ level \citep{HolzwarthR_00a}.

As the pulse circulates in the cavity, dispersion causes a difference
between the group and phase velocities. The pulse envelope, $A(t)$,
propagates at the group velocity, ${\rm v}_{\rm gr}$, while the
carrier wave travels at its phase velocity, causing them to shift with
respect to each other after each pulse round-trip. This is illustrated
in Fig.~\ref{fig:pulse}. Comparing two consecutive pulses with
$A(t)=A(t-T)$, equation (\ref{eq:fourier}) implies that the electric
field is given by
\begin{equation}
  E(t) = E(t-T) e^{-i\,2\pi\nu_{\rm c}T}
       = E(t-T) e^{-i\,2\pi\nu_{\rm ce}T}
\end{equation}
where, in the last step, a large integer multiple of $2 \pi$ was
subtracted from the phase, equation (\ref{eq:fopt}) was used with
$\nu_{\rm c}$ as one of the modes and we recall that $\nu_{\rm
  r}=T^{-1}$. That is, the phase shift for each pulse round-trip is
$\Delta \varphi = 2 \pi \nu_{\rm ce} T$. Therefore, the frequency comb
is offset from being exact integer harmonics of $\nu_{\rm r}$ by the
so-called carrier-envelope offset frequency, $\nu_{\rm ce}=\Delta
\varphi/2\pi T$ \citep{UdemT_02a}. It should be emphasised that the
{\it model} for the origin of $\nu_{\rm ce}$ is not important for
precision measurements using the comb; all that is required is that
$\nu_{\rm ce}$ can be fixed to a certain value experimentally
(e.g.~via a feedback loop) and that the mode spacing, $\nu_{\rm r}$,
is constant.

As Fig.~\ref{fig:pulse} illustrates, the spectral envelope of the
frequency comb has a width that is {\it roughly} given by the inverse
of the pulse duration and a mode spacing given {\it exactly} by the
inverse round-trip time, $\nu_{\rm r}=T^{-1}$. Modern mode-locked
lasers readily achieve $\sim$$100{\rm \,THz}$-wide spectra
($\approx$$830$\,\AA\ wide at $5000$\,\AA) from $\sim$$10{\rm \,fs}$
pulses. Two types of lasers are common: the titanium--sapphire
Kerr-lens mode-locked laser which operates around $8000$\,\AA\ and the
erbium- and ytterbium-doped fibre lasers which operate at around
$1.55$ and $1.04{\rm \,\umu m}$ respectively. Typical values for the
mode spacing/repetition frequency are $\nu_{\rm r}\sim100{\rm \,MHz}$,
but in some special cases up to $\nu_{\rm r}\sim40$--$160{\rm \,GHz}$
has been achieved. See further discussion in Section \ref{sec:prob}.

\subsection{Self-referencing the comb}\label{ssec:self}

\begin{figure}
\centerline{\includegraphics[width=0.9\columnwidth]{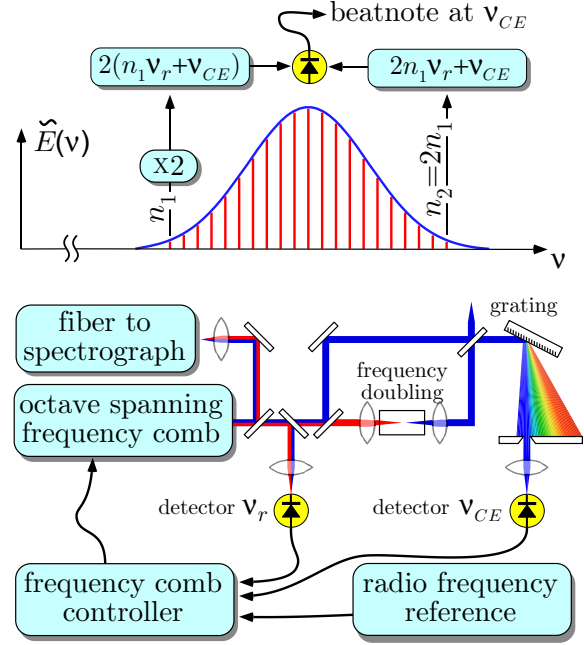}}
\caption{A typical self-referencing scheme. {\it Top}: The carrier
  envelope offset frequency, $\nu_{\rm ce}$, that displaces the modes
  from being exact harmonics of the repetition rate, $\nu_{\rm r}$, is
  measured by frequency doubling some modes at the `red' side of the
  comb and beating them against modes at the `blue' side (depending on
  the type of laser, these do not need to be truly blue or red). {\it
    Bottom}: Such a self referencing scheme employs a dichroic mirror
  to separate the `red' and `blue' wings of the frequency comb. The
  former is frequency-doubled in a non-linear crystal and reunited
  with the `blue' part to create a wealth of beat-frequencies, all at
  $\nu_{\rm ce}$.  }
\label{fig:freqcomb}
\end{figure}

Measuring and/or stabilizing both radio frequencies, $\nu_{\rm ce}$
and $\nu_{\rm r}$, fixes the entire optical frequency comb by virtue
of equation (\ref{eq:fopt}). While the pulse repetition rate,
$\nu_{\rm r}$, is readily measured with a photo-detector somewhere in
the beam, the determination of $\nu_{\rm ce}$ requires significant
effort. In fact it took more than 20 years to develop a simple
$\nu_{\rm ce}$ detection scheme, dubbed `self-referencing', which is
illustrated in Fig.~\ref{fig:freqcomb}. A low frequency mode with mode
number $n_1$ is frequency-doubled to frequency $2(n_1 \nu_{\rm r} +
\nu_{\rm ce})$. This is superimposed with a high frequency mode,
$n_2$, onto a photo-detector. The resulting beat-frequency is
therefore $(2n_1-n_2)\nu_{\rm r} + \nu_{\rm ce}$ which reduces to the
desired signal, $\nu_{\rm ce}$, provided that $2n_1=n_2$. This latter
condition requires the comb to span an entire optical octave. Lasers
capable of producing such broad spectra are only now becoming
available (e.g.~\citealt{MatosL_04a}; \citealt*{FortierT_06a}); they
are currently experimental and have several disadvantages. The
standard method of producing this spectral width has therefore been
via a process called self-phase modulation induced by non-linear
interactions in optical fibres\footnote{Self-phase modulation
  originates from an intensity-dependent refractive index of the
  material. In particular, the polarizability of the material behaves
  non-linearly with respect to the incident electromagnetic field.
  This is also called the `Kerr effect'. It provides an additional
  phase delay proportional to the field's intensity distribution (in
  both space and time).  Applications in the context of the
  calibration comb include intensity-dependent lensing for
  mode-locking (`Kerr-lens mode-locking') in femtosecond lasers and
  spectral broadening or ultra-fast pulse shaping.}. For practical
pulse repetition rates in the radio frequency domain, especially those
approaching the ${\rm GHz}$ regime, the required peak intensity for
such non-linear processes is difficult to achieve because the
available laser power is shared by more pulses. However, with the
development of Kerr-lens mode-locked lasers and photonic crystal
fibres \citep[e.g.][]{KnightJ_96a}, octave-spanning combs can now be
simply generated. The fibre lasers mentioned above use specially doped
fibres to achieve similar spectral broadening.

Self-referencing applies the full accuracy of the radio frequency
reference to the entire optical comb, yielding a perfectly regular
calibration grid for the spectrograph. Current commercial caesium
atomic clocks would provide a calibration as precise as several parts
in $10^{13}$, i.e.~a velocity precision of $\sim$$0.01\,\cms$. This
may even be improved by using a timing signal broadcasted by the
Global Positioning System to further stabilize the caesium clock or by
setting up a local caesium fountain clock \citep[e.g.][]{BauchA_03a}.
However, we demonstrate in Section \ref{sec:phot} that typical echelle
spectrographs will limit the available precision to $\sim$$1\,\cms$
and so such improvements should not be necessary.

In contrast to this standard radio-frequency self-referencing, an
alternative optical referencing scheme might prove more practical for
astronomical applications. In such a scheme the $n$th optical mode,
$\nu_n$, could be stabilized, for example, to an iodine-stabilized
laser. The repetition rate would be stabilized to a precise radio
frequency reference just as in the standard scheme above. As
iodine-stabilized lasers can be as accurate as a few parts in
$10^{13}$ \citep*[e.g.][]{YeJ_01a} this method could also provide the
required accuracy.

\subsection{The comb as a calibration source}\label{ssec:combspec}

Figure \ref{fig:freqcomb} shows the self-referenced (i.e.~frequency
calibrated) comb spectrum being fed into the echelle spectrograph via
an optical fibre. We envisage that such a configuration would offer
the greatest stability and potential for tracking real-time
distortions of the spectrograph wavelength scale. In practice, the
spectrograph would be fed by two fibres, one carrying the science
object light and the other carrying the comb light for calibration.
The two spectra would be recorded simultaneously next to each other on
the CCD. The intensity of the frequency comb spectrum might be varied
according to changes in the intensity of object light being
transmitted by the telescope and spectrograph. Such variations could
easily be tracked via feedback from detection of stray object light as
the main beam passes into the spectrograph. Thus, changes in the
weather conditions or small drifts in the spectrograph itself during
the object exposure, which might ordinarily lead to distortions of the
wavelength scale, would affect the recorded comb spectrum in the same
way.

Of course, for very high precision work, one would have to consider
other instrumental effects, such as differences between (and
inhomogeneities in) the object and comb beam profiles as they emerge
from the fibres. One would also consider placing the entire
spectrograph in vacuum to ensure temperature and pressure stability.
Many of the above techniques have already been demonstrated
\citep[e.g.][]{BaranneA_96a,MayorM_03a}. Of course, the more
traditional approach of taking calibration exposures before and/or
after the object exposure would still be open with the comb as a
calibration source. The comb light could also be expanded and diffused
after emerging from the fibre to illuminate a traditional spectrograph
slit.

Note that the specifics of a final frequency comb design and
spectrograph feed system need careful consideration quite beyond the
scope of this paper. Indeed, these design considerations first require
knowledge of the optimal laser repetition rate for typical echelle
resolutions. The various tolerances involved in the design should also
be informed by the photon-limited calibration precision achievable
with an echelle spectrograph. These quantities are calculated in
Section \ref{sec:phot} and their implications for design challenges
are considered in Section \ref{sec:prob}.

\section{Optimal photon-limited wavelength calibration precision}\label{sec:phot}

The wavelength calibration precision available from a comb spectrum
will clearly improve with increasing line-density. However, since the
spectrograph resolving power is finite, too high a line-density will
decrease the contrast between neighbouring lines and degrade the
available precision; an optimal line-spacing must exist. Here we
calculate this optimum laser repetition rate and the associated
wavelength calibration precision. We first present a simple estimate
of the photon-limited precision using some general arguments in
Section \ref{ssec:approx}. Sections \ref{ssec:info} \& \ref{ssec:res}
describe more rigorous calculations which also provide the optimal
laser repetition rate. These calculations concentrate only on the
photon-limited precision and ignore systematic errors; the latter are
discussed briefly in Section \ref{ssec:err}. We assume throughout this
section that only one CCD exposure of the comb spectrum is used for
wavelength calibrating the spectrograph. The dynamic range of the CCD
pixels therefore limits the comb $S/N$. Clearly, future improvements
in detector technology, such as multiple or continuous reading out of
the comb spectrum during a single science exposure, could improve the
effective comb $S/N$, thereby allowing even more precise calibration.

\subsection{General considerations}\label{ssec:approx}

Some general arguments serve as an intuitive introduction to the
calculations in the following sections. Let us first consider a single
unresolved comb line. When extracted from the CCD, this line will
reflect the instrumental profile of the spectrograph. Assuming this to
be a Gaussian whose FWHM is sampled by $n>1$ pixels, a well-known
approximation to the error on the line's position is
\citep[e.g.][]{BraultJ_87a}
\begin{equation}
\delta v \approx A\frac{\rm FWHM}{S/N \times\!\sqrt{n}}\,.
\end{equation}
The $S/N$ here refers to the peak $S/N$ per pixel across the line
(i.e.~usually at the line centre). The pre-factor, $A$, depends on both
the functional form of the line-profile and the relationship between
$S/N$ and pixel intensity. For example, if the noise level is constant
across the profile then $A=0.693$ for a Gaussian and $0.798$ for a
Lorentzian \citep{BraultJ_87a}. In our case the noise is Poissonian
with an additional small contribution from detector noise. From a
simple numerical experiment in which Gaussian profiles with this noise
profile were fitted with standard least-squares techniques, we find
that $A\approx0.41$ when the photon noise dominates the detector noise.

To determine the total velocity precision available from, say, an
entire echelle order containing such comb lines we must know the line
density and the spectral width of the order. As a first guess, let us
assume that a comb line occurs every 2.5 resolution elements so that
the line-density is high but neighbouring lines do not strongly
overlap.  For consistency with the more rigorous calculations below,
let us assume that echelle orders are 2048 pixels long with 3-pixel
sampling of the FWHM. That is, there are $N\approx272$ comb lines per
echelle order.  Therefore, assuming a peak $S/N=500$ in each line, the
total calibration precision per echelle order at $R=150000$
(i.e.~${\rm FWHM}=2\,\kms$) is $\sigma_v \approx \delta v/\sqrt{N} =
0.41 \times 2\,\kms / (500 \times \!\sqrt{3\times272}) = 5.7\,\cms$.

The following sections demonstrate that our assumption that comb lines
should be spaced by $\sim$2.5 resolution elements does in fact provide
an optimal calibration with precision close to $6\,\cms$ per echelle
order as predicted above.

\subsection{Velocity information content of comb spectra}\label{ssec:info}

If a small\footnote{In all applications discussed in this paper, the
  expected velocity change is much smaller than the width of any
  spectral feature.} velocity change is to be measured from pixel $i$
in a spectrum $F(i)$ with 1-$\sigma$ error array $\sigma_F(i)$, the
limiting velocity precision is given by \citep*{BouchyF_01a}
\begin{equation}\label{eq:sv_i}
\frac{\sigma_{\rm v}(i)}{c} = \frac{\sigma_F(i)}{\lambda(i)\,\left[\partial F(i)/\partial\lambda(i)\right]}\,.
\end{equation}
This equation simply states that a more precise velocity measurement
is available from those pixels where the flux has a large gradient
and/or the uncertainty in the flux is small. We have used wavelength
as the dispersion coordinate here but one may replace $\lambda$ with
any other dispersion measure desired, including pixels. This quantity
can be used as an optimal weight,
\begin{equation}\label{eq:weight_i}
W(i) = \left[\sigma_{\rm v}(i)/c\right]^{-2}\,,
\end{equation}
to derive the total velocity precision available from all pixels in a
spectrum,
\begin{equation}\label{eq:sv}
\frac{\sigma_{\rm v}}{c} = \frac{1}{\sqrt{\sum_i W(i)}}\,.
\end{equation}
If the spectrum is divided into smaller spectral slices $k$, such as
echelle orders, then it follows that the total velocity precision
available is a weighted average of that available from each slice,
\begin{equation}\label{eq:sv_k}
\sigma_{\rm v} = \frac{1}{\sqrt{\sum_k\sigma^{-2}_{\rm v}(k)}}\,.
\end{equation}

\subsection{Simulations of comb echelle spectra}\label{ssec:res}

Equations (\ref{eq:sv}) \& (\ref{eq:sv_k}) can be used to calculate
the velocity precision achievable with a frequency comb recorded with
an echelle spectrograph. To this end we construct a simple model of a
frequency comb spectrum by writing the flux in each pixel, measured in
photo-electrons, as the product of a slowly varying `envelope
function' $F_{\rm e}(i)$ and a quickly varying `comb function' $f(i)$
which varies between zero and unity: $F(i)=f(i)F_{\rm e}(i)$. A
reasonable model of the expected 1-$\sigma$ error array is then
\begin{equation}
\sigma_f(i) = \sqrt{f(i)F_{\rm e}(i)+\sigma^2_{\rm d}(i)}\,,
\end{equation}
where $\sigma_{\rm d}(i)$ is the noise contribution from the detector
in each pixel. The signal-to-noise ratio for each pixel can then be
written as
\begin{equation}
\left(\frac{S}{N}\right)(i) = f(i)\sqrt{\frac{F_{\rm e}(i)}{f(i)+\beta^2_{\rm d}(i)}}\,,
\end{equation}
where $\beta_{\rm d}(i) \equiv \sigma_{\rm d}(i)/\sqrt{F_{\rm e}(i)}$
is the ratio of the detector noise and photon noise expected from the
envelope at pixel $i$. Assuming that $\sigma_{\rm d}(i)=\sigma_{\rm
  d}$ is constant for all pixels and recalling that $F_{\rm e}(i)$
varies slowly compared to $f(i)$ then the maximum $S/N$ of the comb is
\begin{equation}
\left(\frac{S}{N}\right)_{\rm max} = \sqrt{\frac{F_{\rm e,max}}{1+\beta^2_{\rm d,max}}}\,.
\end{equation}
Therefore, in our simulations one specifies the signal-to-noise at all
pixels by knowing the shape of the comb and envelope functions and by
specifying the maximum signal-to-noise, $(S/N)_{\rm max}$, across the
comb and a constant detector noise, $\sigma_{\rm d}$ or $\beta_{\rm d,max}$.

\begin{figure}
\centerline{\includegraphics[width=\columnwidth]{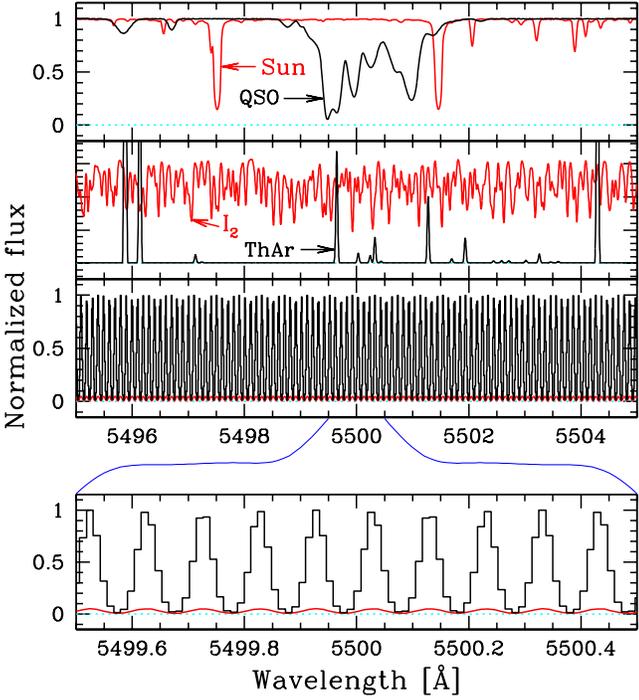}}
\caption{Comparison of small sections of simulated object spectra with
  different calibration spectra. The upper panel shows a model Solar
  spectrum (red/grey line) and a low-ionization metal transition
  (arbitrary redshift) with typical velocity structure in a synthetic
  quasar absorption spectrum (black line). The next panel compares
  model spectra of an iodine absorption cell (red/grey line) and ThAr
  emission-line lamp (black line) at $R=150000$ on different arbitrary
  flux scales.  Note the complexity of the I$_2$ absorption spectrum
  and the relatively sparse distribution of ThAr lines with their
  large range of intensities. The third panel shows a simulated
  frequency comb echelle spectrum over the same wavelength region with
  repetition rate $\nu_{\rm r}=10{\rm \,GHz}$, resolving power
  $R=150000$ and $(S/N)_{\rm max}=500$. The lower panel zooms in on a
  1-\AA\ wide section of the upper panel.  For the comb spectra, the
  black histogram shows the expected flux while the red/grey solid
  line shows the error array exaggerated by a factor of 25.}
\label{fig:comb_spec}
\end{figure}

For our simulations we assume that the envelope function is constant
over the entire wavelength range of the comb. That is, all comb lines
have equal intensity. The comb function can be modelled as a series of
delta functions equally spaced in frequency according to the assumed
laser repetition rate, $\nu_{\rm r}$. The delta-functions are
convolved with the spectrograph's instrumental profile which we assume
to be a Gaussian of width specified by the resolving power $R$.  For
an echelle spectrograph $R$ is roughly constant over the entire
wavelength range and we make this additional assumption here.
Simulated spectra are placed on a log-linear wavelength scale sampled
at 3 pixels per resolution element (FWHM) in order to emulate
one-dimensional reduced echelle spectra. A portion of simulated
spectrum is shown in Fig.~\ref{fig:comb_spec}.

We calculate $\sigma_{\rm v}$ from the simulated spectra using equation
(\ref{eq:sv}). In calculating the derivative in equation
(\ref{eq:sv_i}) we have simply used the finite-difference derivative
between successive pixels,
\begin{equation}\label{eq:finite}
\frac{\partial F(i)}{\partial\lambda(i)} \approx
\frac{F(i)-F(i-1)}{\lambda(i)-\lambda(i-1)}\,.
\end{equation}
With this application, the first pixel of any spectral subdivision has
undefined weight, $W(i)$. The sum in equation (\ref{eq:sv}) therefore
begins at the second pixel in each spectral subdivision.

\begin{figure}
\centerline{\includegraphics[width=\columnwidth]{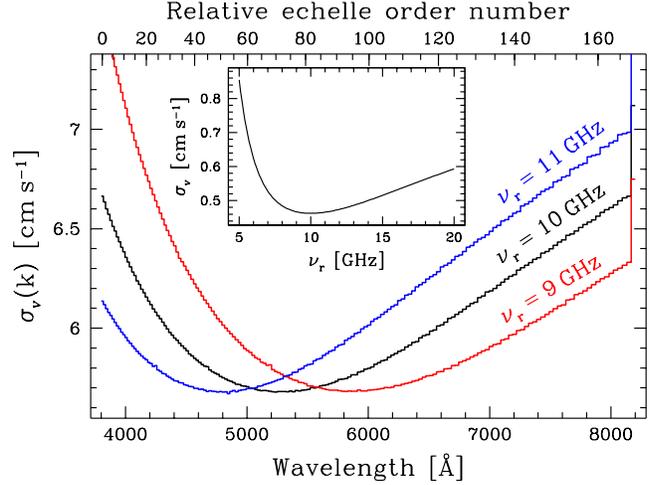}}
\caption{Results from comb simulations with $R=150000$, $(S/N)_{\rm
    max}=500$ and $\beta_{\rm d,max}=0.01$. The spectrograph's
  resolution (FWHM) is sampled with 3 pixels and each echelle order is
  assumed to be 2048 pixels long.  The main panel shows the expected
  velocity precision available from each echelle order, $\sigma_{\rm
    v}(k)$, for three different line-spacings while the inset shows
  the total velocity precision, $\sigma_{\rm v}$, integrated over the
  full wavelength range of the simulated spectrum,
  $3800$--$8200$\,\AA. For a comb spectrum of this resolution the
  optimum line-spacing (or repetition rate) is $\nu_{\rm r}=10{\rm
    \,GHz}$, resulting in a total velocity uncertainty of $\sigma_{\rm
    v}=0.45\,\cms$. $\sigma_{\rm v}$ rises sharply for repetition
  rates below the spectrograph's resolution (3.3\,GHz at 6000\,\AA).}
\label{fig:sv_vs_dnu}
\end{figure}

An example of the results from the simulated comb spectra is shown in
Fig.~\ref{fig:sv_vs_dnu}. Here we assumed $R=150000$ and $(S/N)_{\rm
  max}$ of 500\,per spectral pixel. Gemini/bHROS, ESO-3.6m/HARPS and
Subaru/HDS already operate at (or close to) this resolving power. To
obtain $(S/N)_{\rm max}=500$\,per pixel with modern CCDs, which
typically have a linear response when fewer than $\sim$$40000$
photo-electrons are detected per pixel, the comb spectrum would have
to be spread over 6--7 pixels in the spatial direction; this is
readily achievable with most modern echelle spectrographs. We also
assumed a fairly typical wavelength range, $3800$--$8200$\,\AA.

Fig.~\ref{fig:sv_vs_dnu} demonstrates two important results: (i) The
velocity precision available from each echelle order is
$\sim$$6$--$7\,\cms$ which is consistent with the
`back-of-the-envelope' calculation in Section \ref{ssec:approx}. This
varies by $\la$20\,per cent across all echelle orders because the comb
lines are uniformly separated in frequency space; (ii) The total
velocity precision integrated over all echelle orders is $0.45\,\cms$
but, for a given $R$, this depends strongly on the comb line-spacing.
Assuming that each spectral resolution element is sampled at or above
the Nyquist rate and the detector noise is small compared to the
photon noise, the optimum line-spacing required is determined by the
resolving power:
\begin{equation}\label{eq:rep}
\nu^{\rm opt}_{\rm r} \approx \frac{1.5\times10^6}{R}{\rm \,GHz} \sim 3\Delta\nu_{\rm cent}\,,
\end{equation}
where $\Delta\nu_{\rm cent}$ is the spectrograph's resolution in
frequency space at approximately the centre of its wavelength range.
Also, at constant $R$ and $(S/N)_{\rm max}$ (and again assuming
$\beta_{\rm d,max}\ll 1$), the increase in overall velocity
uncertainty with increasing $\nu_{\rm r}$ is simply described as
$\sigma_{\rm v} \propto \sqrt{\nu_{\rm r}}$ provided that comb lines
are well resolved from each other, i.e.~$\nu_{\rm r}\ga1.5\,\nu^{\rm
  opt}_{\rm r}$. The increase in $\sigma_{\rm v}$ with decreasing
line-spacing below $\nu^{\rm opt}_{\rm r}$ is extremely fast but has
no simple analytic description. Suffice it to say that our simulations
show that the increase is faster than exponential. Nevertheless, a
total precision within a factor of 2 of the optimum can be achieved
over the range $\nu_{\rm r}\approx(0.5$--$5)\,\nu^{\rm opt}_{\rm r}$.
Equivalently, a comb of fixed $\nu_{\rm r}$ can be used to
almost-optimally calibrate spectra over a reasonable range in
resolving power.

These results are easily scaled according to the parameters of
different spectrographs. It is not surprising to find that
$\sigma_{\rm v} \propto 1/(S/N)_{\rm max}$ and $\sigma_{\rm v} \propto
1/R$ if, again, the sampling of each resolution element is adequate
and $\beta_{\rm d,max}\ll 1$.  Thus, for a given resolving power and
maximum $S/N$, the optimum photon-limited precision available over the
wavelength range $3800$--$8200$\,\AA\ is
\begin{eqnarray}
\sigma_{\rm v}^{\rm opt}\!\!\!\! & = & \!\!\!\!0.45\,\left[\frac{500}{(S/N)_{\rm max}}\right]\left(\frac{1.5\times10^5}{R}\right)\sqrt{\frac{\nu^{\rm opt}_{\rm r}}{10{\rm \,GHz}}}\,\cms\label{eq:svopt1}\\
\!\!\!\! & = & \!\!\!\!0.45\,\left[\frac{500}{(S/N)_{\rm max}}\right]\left(\frac{1.5\times10^5}{R}\right)^{3/2}\,\cms\label{eq:svopt2}\,.
\end{eqnarray}
For example, with lower resolving powers typical of spectrographs like
VLT/UVES and Keck/HIRES ($R=70000$), and assuming that $(S/N)_{\rm
  max}=300$ can be easily obtained in a single comb exposure, a
precision of $\sigma_{\rm v}^{\rm opt}=2.4\,\cms$ should be
achievable. The pre-factor of $0.45\,\cms$ in equations
(\ref{eq:svopt1}) \& (\ref{eq:svopt2}) is larger for narrower
wavelength ranges. For example, it is $0.63\,\cms$ for
$4500$--$6500$\,\AA.

\subsection{Additional sources of error}\label{ssec:err}

The typical photon-limited precision of $\sim$$1\,\cms$ calculated
above is 5 orders of magnitude smaller than the typical pixel size for
most modern echelle spectrographs. To provide some context, for a
15-$\umu$m pixel this corresponds approximately to the size of the
silicon atoms making up the CCD substrate! Therefore, if the photon
noise limit is to be reached, mitigation of possible systematic
effects in the telescope, spectrograph and detector systems is
imperative. Many such effects have already been considered for
observations requiring short-term stability ($\sim$$1$ night to
$\sim$$1$ week) which achieve $100$-\cms\ precision or better
\citep[e.g.][]{LovisC_06a}.  However, in pushing to the $1$-\cms\
precision level, additional sources of error will need to be
considered in detail.

We will not conduct an exhaustive study of possible effects here but
the following initial list is illustrative: (i) Inhomogeneities and
variations in the intensity profile of the object and comb fibres
presented at the spectrograph slit; (ii) Variations in the
spectrograph's instrumental profile (IP) across the CCD; (iii)
Inaccuracies in the echelle grating's line markings; (iv) Variations
in pixel size and separation across an individual CCD array; (v)
Intra-pixel sensitivity variations; and (vi) CCD temperature
variations, especially during read-out. While pre-slit effects such as
(i) clearly require their own solutions (in this case, efficient beam
homogenization and stabilization), most post-slit effects such as
(ii--vi) might be mitigated by using the frequency comb itself. Since
the comb provides a series of closely spaced modes which are known,
{\it a priori}, to have equidistant spacing, comb spectra could be
used to accurately characterize the spectrograph and detector's
response to different systematic effects. In practice, this would be
facilitated by injecting the comb spectrum into the object fibre and
by altering $\nu_{\rm r}$ and/or $\nu_{\rm ce}$ slightly to shift the
comb spectrum by less than a resolution element. Thus, while the
frequency comb would provide, in principle, a significant increase in
precision, it would also provide an important means to characterize,
measure and correct-for many systematic errors so that the available
photon-limited precision could be realised in practice.

Aside from the spectrograph's IP, an additional factor contributing to
asymmetries and variations in the measured point-spread function (PSF)
could be intrinsic asymmetries in the comb modes themselves. Although
the central frequencies of the modes are defined precisely by the
self-referencing system, asymmetries in the modes' profiles could
arise via correlated phase and amplitude noise, generally referred to
as `repetition-rate noise'. The time-dependence (and therefore the
time-averaging properties) of the asymmetry in individual modes will
depend greatly on the type of laser comb system eventually employed.
Nevertheless, since the spectral width of individual modes is
typically $\sim$1\,MHz (0.6\,\ms\ at 6000\,\AA) and the asymmetry is
expected to be small in comb systems stabilized to precise clocks, the
difference between an individual mode's frequency and the centroid of
its intrinsic profile should amount to no more than $\sim$1--10\,\cms.
If the asymmetries vary very slowly with time, this effect would be
corrected for when modelling the overall PSF. However, if mode
profiles were found to vary on very short time-scales, additional
techniques for reducing the mode line-width closer to the desired
calibration uncertainty (i.e.~$\sim$1\,\cms\ or 17\,kHz at 6000\,\AA)
might be employed \citep[e.g.][]{DiddamsS_03a}.

\section{Design challenges and possible solutions}\label{sec:prob}

From the point of view of designing a real frequency comb calibration
system, the most important result from the simulations in Section
\ref{ssec:res} is that laser repetition rates of $\nu_{\rm
  r}=5$--$30{\rm \,GHz}$ give the highest velocity precision for
practical spectrograph resolving powers of $R=50000$--$300000$
[equation (\ref{eq:rep})]. Although high repetition rate ($>10{\rm
  \,GHz}$) lasers with picosecond pulses and frequency combs with
broad spectral coverage, sub-100-fs pulses and low repetition rate
($<1{\rm \,GHz}$) have been demonstrated, the desired combination of
high repetition rates, short pulses and large spectral widths
(e.g.~$\sim$3800--8200\,\AA) is challenging and is yet to be
experimentally demonstrated. A uniform intensity distribution across
the whole wavelength range is also highly desirable since this ensures
that similar velocity precision can be obtained from any spectral
sub-division. Below we discuss why these design requirements may be
problematic and offer some possible solutions.

\subsection{Repetition rate}\label{ssec:rep}

\begin{figure}
\centerline{\includegraphics[bb = 65 0 556 303,width=\columnwidth]{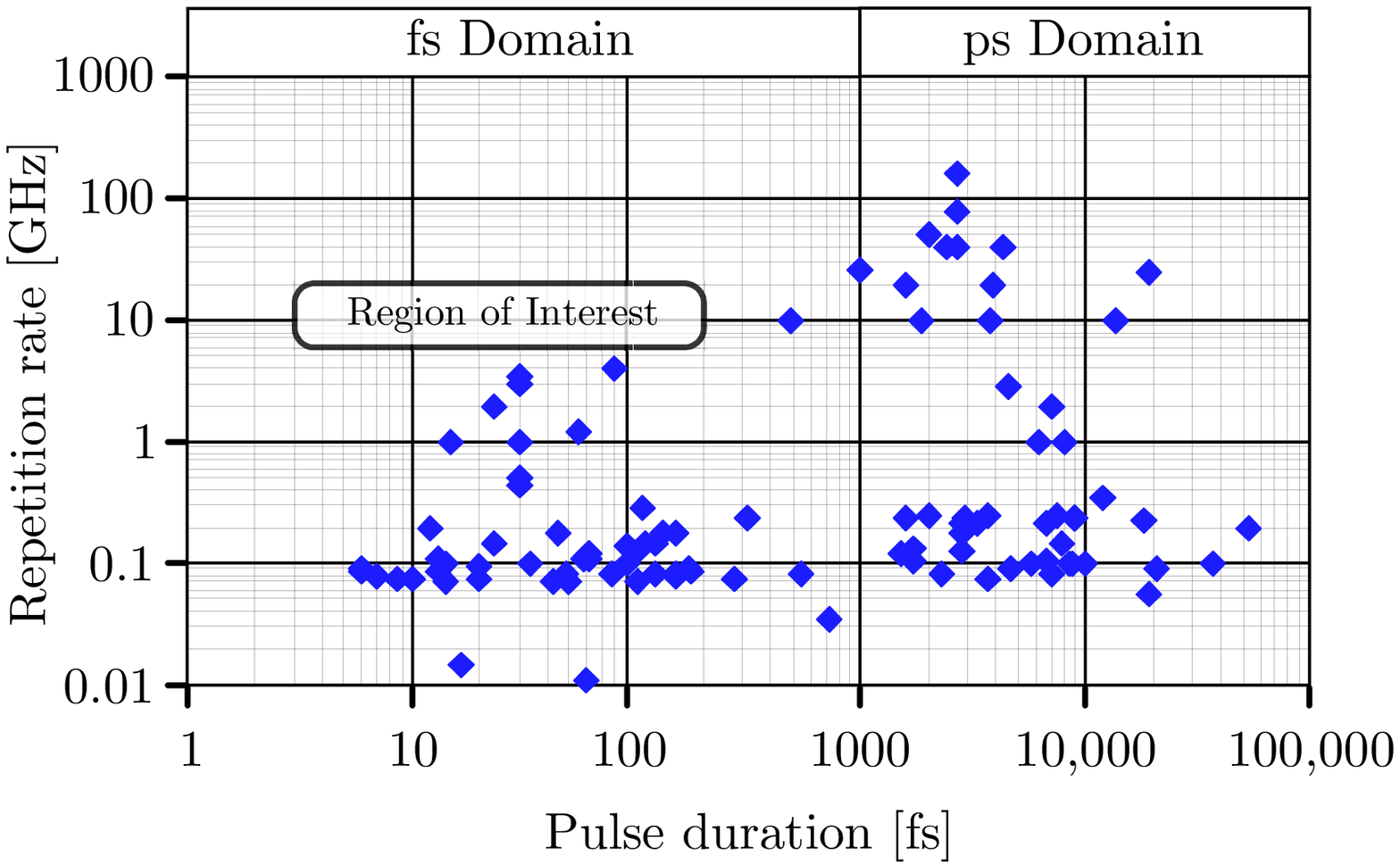}}
\caption{A review of femtosecond laser performance data shows two
  distinct clusters, one in the 10--200-fs regime and the other in the
  1--20-ps regime of pulse durations. Sub-100-fs pulse lasers seem to
  operate to up to 4 GHz. The region of interest (5--30\,GHz
  repetition rate, 3--200\,fs pulse duration) is still new territory
  for femtosecond pulsed lasers.}
\label{fig:lasers}
\end{figure}

Reaching the optimum repetition rate suggested by
Fig.~\ref{fig:sv_vs_dnu} is not, by itself, challenging; several
systems with even higher $\nu_{\rm r}$ already exist. For example,
\citet{LecomteS_05a} published a mode-locked laser with $\nu_{\rm
  r}=40{\rm \,GHz}$. Unfortunately, the peak power of such lasers is
so low that they cannot use the fast but weak pulse shaping and
shortening mechanisms that are usually provided by the Kerr effect.
Instead these lasers rely on slow saturable absorbers for mode
locking, forcing pulse durations into the picosecond regime with the
associated narrow spectral widths. Indeed, the spectral width would
need to be $\ga$30 times larger to maintain the comb's coherence in
the spectral broadening process discussed below in Section
\ref{ssec:band}. The challenge of generating sub-100\,fs pulses at
repetition rates $\ga$10\,GHz is illustrated by a review of the
performance of ultra-fast lasers in the literature, as summarised in
Fig.~\ref{fig:lasers}. To our knowledge, the highest repetition rate
achieved so far in the sub-100-fs regime is $\nu_{\rm r}\approx4{\rm
  \,GHz}$ \citep{LeburnC_04a}.

One possible way to combine high repetition rate, short pulses and
large spectral width is to spectrally filter the laser beam outside
the cavity. The strategy would be to use a laser with modest
repetition rate, say $\nu_{\rm r}\sim2{\rm \,GHz}$, where reasonably
short pulses can still be generated and where there is still enough
energy per pulse to allow efficient spectral broadening (as discussed
in Section \ref{ssec:band}). The effective mode spacing could then be
increased by filtering out most modes with a Fabry-Perot etalon with a
free spectral range chosen to be an integer multiple of $\nu_{\rm r}$
and equal to the final desired mode spacing \citep[e.g.][]{UdemT_99a}.
The etalon would be required to have highly stable transmission
fringes and a high enough finesse to ensure strong suppression of the
extraneous modes.

\begin{figure}
\centerline{\includegraphics[width=0.9\columnwidth]{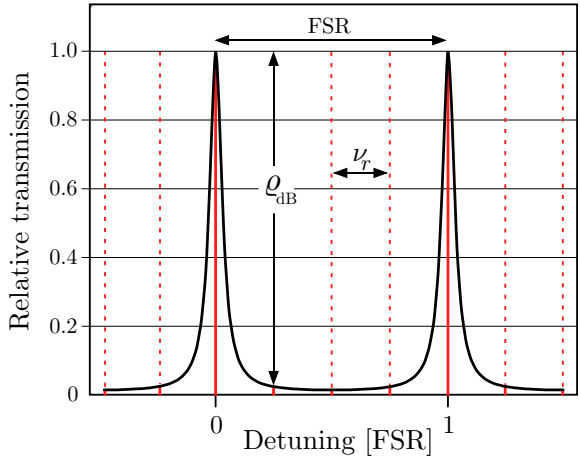}}
\caption{Simple representation of the (Fabry-Perot) transmission
  profile of an idealized mode-filter cavity (black curves). The
  resulting mode spacing is given by the free spectral range (FSR)
  between transmission peaks. The degree to which the intervening modes
  are suppressed is specified by the side-mode suppression ratio,
  $\rho_{\rm dB}$, measured at the position of the strongest
  off-resonant mode. Achieving strong side-mode suppression requires a
  high finesse according to equation (\ref{eq:finesse}).}

\label{fig:filter}
\end{figure}

Figure \ref{fig:filter} shows an etalon transmission function defined
by the free spectral range (FSR), mode suppression ratio in decibels
($\rho_{\rm dB}$) and by a repetition rate enhancement factor
$q\equiv{\rm FSR}/\nu_{\rm r}$. Since the cavity transmission profile
should be adequately described by an Airy function, the required
finesse of the etalon cavity is uniquely determined by $\rho_{\rm dB}$
and $q$:
\begin{equation}\label{eq:finesse}
F=\frac{\pi}{2} \frac{\sqrt{10^{\rho_{\rm dB}/10}-1}}{\sin(\pi/q)} \approx \frac{q}{2}10^{\,\rho_{\rm dB}/20}\,,
\end{equation}
where the approximation is accurate for suppression ratios $>10{\rm
  \,dB}$ and $q\gg1$. For example, the approximation is correct within
1.6 per cent for $\rho_{\rm dB}=50{\rm \,dB}$ and $q=10$. For a 10-GHz
FSR, a side-mode suppression of $\rho_{\rm dB}=50{\rm \,dB}$ can be
achieved by eliminating 3 out of every 4 modes (i.e.~$q=4$) from a
$\nu_{\rm r}\sim2.5{\rm \,GHz}$ comb, requiring a finesse of
$F\approx700$. Such a repetition rate and finesse are achievable with
current technology and so this mode-filter technique seems a promising
approach to achieve higher effective repetition rates. However, one
challenge will be for a large spectral range to be filtered in this
way because the dispersion of the cavity components will create
differing path-lengths for light of widely different wavelengths. The
etalon mode-spacing remains constant only if the dispersion is
compensated properly. Simultaneously achieving reliable dispersion
compensation and high finesse/reflectivity may impose a challenge for
the etalon mirror design \citep{GohleC_05a}. A practical solution
might employ several frequency combs and etalons which each cover a
narrower spectral range.

\subsection{Spectral width}\label{ssec:band}

In most octave-spanning frequency combs, the process of self-phase
modulation is used to extend the comb's spectral width. For example,
titanium--sapphire femtosecond laser pulses launched into photonic
crystal fibres or micro-structured fibres have shown extensive
spectral broadening and this has been used for generating frequency
combs with $\nu_{\rm r}$ up to 1\,GHz
\citep[e.g.][]{JonesD_00a,HolzwarthR_00a}. In this process the
spectral coverage is increased fairly symmetrically around the central
wavelength of the pulses and a total spectral coverage from
$4500$\,\AA\ to $1.4{\rm \,\umu m}$ has been achieved. Similarly,
lasers based on Er-doped fibres emitting around $1.5{\rm \,\umu m}$
have produced spectra covering 5000\,\AA\ to $2.2{\rm \,\umu m}$ at
$\nu_{\rm r}$ up to 250 MHz \citep[e.g.][]{NicholsonJ_03a,MackowiakV_05a}.

However, producing broad spectra at higher repetition rates is
difficult because the self-phase modulation process requires a certain
pulse peak power to provide an octave of broadening.  Assuming a pulse
train at $\nu_{\rm r}$ with an average power, $\left<P\right>$, the
peak power is roughly given by the energy in one pulse,
$\left<P\right>/\nu_{\rm r}$, divided by the pulse duration.
Typically, a pulse energy of 2\,nJ for 100-fs pulses is needed,
translating to average powers of 2 \& 20\,W for 1 \& 10\,GHz
repetition rates, respectively. In practice, $\left<P\right>$ in
mode-locked lasers have so far been limited to $\sim$$1{\rm \,W}$.
Thus, the highest repetition rates achieved so far for octave-spanning
combs are around $\nu_{\rm r}\sim1{\rm \,GHz}$. The fact that
sufficient self-phase modulation requires a certain pulse peak power
means that, in practice, broad spectra can be produced only with
relatively low $\nu_{\rm r}$ while achieving higher $\nu_{\rm r}$
reduces the peak pulse power below that required for sufficient
broadening.  For example, the repetition rate of 4\,GHz demonstrated
by \citet{LeburnC_04a} is already close to optimal for very high
spectrograph resolutions $R\sim300000$ [equation (\ref{eq:rep})]. With
a modest effort to filter the modes with an etalon, the effective
repetition rate could probably be increased to $\ga$$10\,{\rm GHz}$.
However, this laser emits between $1.40$ and $1.58{\rm \,\umu m}$ so
significant non-linear conversion processes would be required, such as
frequency doubling (or tripling), or spectral shifting and broadening;
the pulse peak power is too low at such high repetition rates for
these processes to be effective.

On the other hand, a new class of mode-locked lasers, optimized for
larger spectral width rather than shorter pulses, is currently under
development. Such systems have reached spectral widths extending from
$5700$\,\AA\ to $1.3{\rm \,\umu m}$ \citep{MuckeO_05a} at repetition
rates up to $1{\rm \,GHz}$. However, even these broad spectra are not
yet sufficient to cover the entire optical region, especially the blue
region of the spectrum. The current limitation is actually due to the
laser cavity mirrors which do not operate efficiently outside this
(already relatively broad) range.

Another option to achieve the desired effective spectral width could
be the use of very fast and efficient optical phase modulators. These
so-called `optical frequency comb generators'
\citep*[e.g.][]{KourogiM_93a,ImaiK_98a} are capable of adding several
hundred modulation side-bands to the spectrum of a single-mode laser
with separations at the required optimal mode spacing
(i.e.~$\sim$10\,GHz).  However, so far the spectral width of these
comb generators is insufficient for covering the whole optical range.
Therefore, dozens (even hundreds) of generators would have to be used
simultaneously or some advanced scanning method would have to be
devised which allowed the full calibration region to be swept with a
single generator (or perhaps many) within the exposure time.

In summary, designing a frequency comb covering the entire optical
range while maintaining a high repetition rate would require some
improvements in comb technology. A practical solution might employ a
combination of mode filtering stages and nonlinear conversion stages,
possibly with additional amplifier stages to keep the power
sufficiently high.

\subsection{Uniform intensity}\label{ssec:inten}

Unlike the sketch in Fig.~\ref{fig:pulse}, real frequency combs
display strong intensity structure; intensity variations of 2 orders
of magnitude are to be expected, even over relatively short wavelength
ranges ($\sim$$300$\,\AA), especially when using photonic crystal fibres
for spectral broadening \citep[e.g.][]{HolzwarthR_01a}. The intensity
structure is also sensitive to the polarization and peak power of the
pulses coupled into the non-linear fibre. They therefore vary with
time as the laser--fibre coupling and laser parameters change.
Fortunately, since only a small amount of power is required to fully
expose the CCD (amounting to $\sim$$40000$ photons per CCD pixel over
$\sim$10--3600-s exposure times), the spectrum can be flattened by an
attenuation mechanism which can respond to both spectral and time
variations. One such mechanism, for example, might be a spatial light
modulator \citep[e.g.][]{WeinerA_00a} whereby groups of modes
(i.e.~small chunks of the comb spectrum) can be separated,
differentially attenuated and recombined again. An alternative
approach is to improve the matching of phase velocities (`phase
matching') between the various modes in the non-linear broadening
process. This can be achieved by tailoring the longitudinal
dispersion properties of the fibre and has been shown both
theoretically and experimentally to produce broad, flat spectra
\citep{HoriT_04a}.

\section{Conclusion}\label{sec:conc}

We have outlined the possible use of frequency combs as calibration
sources in astronomical echelle spectroscopy. Such combs generate a
series of equally spaced, very narrow modes which, by employing
self-referencing (or possibly other calibration) techniques, have
absolute frequencies which are known {\it a priori} to relative
precisions better than $10^{-12}$. These are significant advantages
compared to the properties of other calibration sources such as ThAr
emission-line lamps and I$_2$ absorption cells. Simulations of comb
spectra with uniform intensity over the optical range were sampled
according to the properties of typical high resolution echelle
spectrographs and CCD detectors. These revealed that the laser
repetition rate must be $\nu_{\rm r}\sim 10$--$20{\rm \,GHz}$ to
produce optimal line-spacings for $R\sim50000$--$200000$
spectrographs. For such systems, the photon-limited wavelength
calibration precision is in the \cms\ regime when integrated over
$\sim$$1000$\,\AA\ ranges and is generally $\la$10\,\cms\ when
integrated over individual echelle orders. The regular grid of comb
lines would also greatly facilitate the tracking and effective removal
of systematic distortions of the wavelength scale which might be
produced through a variety of different effects in the telescope,
spectrograph and detector systems. The absolute wavelength calibration
would therefore be stable over very long time-scales even if elements
of the spectrograph or comb changed.  Indeed, frequency comb
calibration would place spectra observed on different telescopes at
different times on the same absolute, high-precision wavelength scale.

However, the results of the simulations reveal several challenges in
designing and implementing a working comb calibration system. Firstly,
stable self-referenced combs with repetition rates of $\nu_{\rm r}\sim
10$--$20{\rm \,GHz}$ and which cover the entire optical range have not
yet been realised. Another significant challenge will be to ensure
that the comb envelope function -- the peak intensity of the comb
lines -- is uniform over the wavelength range of interest.
Nevertheless, frequency comb technology is improving at a rapid pace
and we expect that these challenges might be overcome by a relatively
small research and development effort. The potential gain is clear:
frequency comb calibration may effectively remove wavelength
calibration uncertainties from all practical high-resolution
spectroscopy and may allow the reliable combination of data taken on
different telescopes over many decades.

\section*{Acknowledgments}

We are indebted to Savely Karshenboim for first suggesting the
possibility of using frequency combs for wavelength calibration of
astronomical spectra and to Robert Wynands for many early discussions
about the practicality of such a system. We thank Victor Flambaum and
Jochen Liske for many informative and helpful discussions. MTM thanks
STFC for an Advanced Fellowship at the IoA.


\bspsmall

\label{lastpage}

\end{document}